\documentclass[pra,twocolumn,nofootinbib,floatfix,10pt]{revtex4-2}

\usepackage[T1]{fontenc}
\usepackage{amsmath}
\usepackage{amssymb}
\usepackage{wasysym}
\usepackage{graphicx}
\usepackage{color,soul}
\definecolor{revblue}{rgb}{0,0,0}
\usepackage{physics}
\usepackage{siunitx}
\usepackage{dsfont}
\usepackage{float}
\usepackage[english]{babel}
\usepackage{blindtext}
\usepackage[english,nomargin,inline,marginclue,draft]{fixme}
\pdfpageheight\paperheight
\pdfpagewidth\paperwidth

\usepackage[colorlinks,linkcolor=blue,anchorcolor=blue,citecolor=blue,urlcolor=blue]{hyperref}

\fxusetheme{colorsig}
\FXRegisterAuthor{cg}{acg}{CG}  
\FXRegisterAuthor{th}{ath}{\color{blue}TH}  
\FXRegisterAuthor{ib}{aib}{\color{red}IB} 
\FXRegisterAuthor{sh}{ash}{\color{cyan}SH} 
\FXRegisterAuthor{db}{adb}{\color{green}DB} 
\FXRegisterAuthor{ps}{aps}{PS}
\makeatletter
\renewcommand*\FXLayoutInline[3]{%
  {\@fxuseface{inline}\ignorespaces{\color{fx#1}[#3: #2]}}}
\makeatother

\long\def\symbolfootnote[#1]#2{\begingroup%
\def\thefootnote{\fnsymbol{footnote}}\footnotetext[#1]{#2}\endgroup}

\def\nobreakbefore{%
  \relax\ifvmode\else
    \ifhmode
      \ifdim\lastskip > 0pt\relax
        \unskip\nobreakspace
      \else 
        \nobreakspace
      \fi
    \fi
  \fi
}
\let\oldcite\cite
\renewcommand\cite{\nobreakbefore\oldcite}

\begin{document}
\title{Ultra-broadband Anti-Jamming Communication via a Rydberg Atomic Receiver}

\author{Jia-Dou Nan$^{1,2,\textcolor{blue}{\star}}$}
\author{Jun-Rong Chen$^{3,4,\textcolor{blue}{\star}}$}
\author{Bang Liu$^{1,2}$}
\author{Qi-Feng Wang$^{1,2}$}
\author{Yu Ma$^{1,2}$}
\author{Yi-Ming Yin$^{1,2}$}
\author{Tian-Yu Han$^{1,2}$}
\author{Guang-Can Guo$^{1,2}$}
\author{Hao Tian$^{3}$}
\author{Li-Hua Zhang$^{1,2,\textcolor{blue}{\S}}$}
\author{Bo Du$^{4}$}
\author{Bin-Bin Wei$^{4,5,\textcolor{blue}{\ddagger}}$}
\author{Dong-Sheng Ding$^{1,2,\textcolor{blue}{\dagger}}$}
\author{Bao-Sen Shi$^{1,2}$}

\affiliation{$^1$Key Laboratory of Quantum Information, University of Science and Technology of China; Hefei, Anhui 230026, China.}
\affiliation{$^2$Anhui Province Key Laboratory of Quantum Network, University of Science and Technology of China, Hefei 230026, China.}
\affiliation{$^3$School of physics, Harbin Institute of Technology, Harbin, Heilongjiang 150001, China.}
\affiliation{$^4$Qianyuan Laboratory, Hangzhou, Zhejiang 310024, China.}
\affiliation{$^5$Institute of system engineering, Tianjin 300161, China.}

\date{\today}

\symbolfootnote[1]{J.D.N. and J.R.C. contribute equally to this work.}

\symbolfootnote[4]{zlhphys@ustc.edu.cn}
\symbolfootnote[3]{weibb.2009@tsinghua.org.cn}
\symbolfootnote[2]{dds@ustc.edu.cn}

\maketitle

\textbf{Ultra-broadband anti-jamming communication represents a promising approach to secure and robust information transfer through {\color{black}spread-spectrum techniques}, effectively combatting malicious interference and eavesdropping. Rydberg atoms, enhanced by waveguide coupling, facilitate ultra-broadband spectrum sensing without traditional RF components. This framework provides an experimental platform for ultra-wide anti-jamming communication. Here, we demonstrate real-time signal demodulation based on frequency-hopping spread spectrum (FHSS) in a waveguide-coupled Rydberg receiver, achieving ultra-broad frequency-hopping covering 100 kHz to 20 GHz and a hopping rate of 100 khop/s. {\color{revblue}When confined to a standard operational band (e.g., the 2.4 GHz ISM band), our system achieves a high channel density of $\sim$8 channels per MHz. Beyond this, by leveraging its ultra-broad and continuous bandwidth, the system supports over 150,000 channels.} 
Experimental results reveal a 51 dB enhancement in narrowband interference tolerance compared with single-frequency systems, confirming its outstanding anti-jamming capability. The reported system demonstrates significant potential for secure communications based on quantum technology, especially communication in complex electromagnetic environments.}

\section*{Introduction}

    {\color{revblue}The rapid growth of wireless communications creates unprecedented demands for enhanced data transmission security and spectral efficiency, leading to severely congested spectrum resources and increasingly stringent requirements for both civilian and military applications. Conventional spectrum-spread techniques can mitigate interference in civilian bands\cite{torrieri2005principles}. Bluetooth, for instance, employs frequency-hopping spread spectrum (FHSS) in the 2.4 GHz ISM band\cite{challoo2012overview, ali2021systematic}.} As the most widely deployed spectrum-spread technique, FHSS enhances the anti-jamming and security capabilities of communication systems by rapidly switching the carrier frequency across a predefined set of channels \cite{qiu2022improved, chen2025ambiguity}. It finds wide applications in military wireless communication \cite{montgomery1988large}, advanced extremely high-frequency satellites \cite{aykin2018adaptive, fritz1999modeling}, emerging mobile systems \cite{hu2007cognitive}, etc. However, conventional FHSS communication systems require extensive hardware support and complex demodulation algorithms to maintain a high performance across the broad operational bandwidths, fundamentally limiting their scalability \cite{liu2020ultrafast}. 
    {\color{revblue}Novel anti-jamming technologies with simple architectures and ultra-broadband operating ranges based on new physical dimensions are being developed to address specific limitations in existing communication frameworks. These approaches offer the potential to extend the capabilities of current systems and provide robust performance under conditions where conventional methods may be constrained.}

    {\color{black} Compared to conventional FHSS receivers that rely on antenna-based front-ends coupled with discrete, bandwidth-limited electronic mixers, receivers based on Rydberg atoms (Rydberg receivers) offer a paradigm shift by integrating the sensing and down-conversion functions into the atomic medium itself. This intrinsic mixing capability enables direct, wideband heterodyne reception, yielding distinct system-level advantages including exceptional sensitivity, ultra-broad continuous frequency response, and lower local oscillator (LO) power requirements.} Rydberg atoms, referring to atoms in highly excited states, have gained attention as a promising candidate for high-sensitivity MW sensing owing to their large electric dipole moment \cite{zhang2024rydberg, Schlossberger2024, yuan2023quantum, liu2023electric, adams2019rydberg}. Precise measurement of MW signals utilizing Rydberg atoms can be achieved based on electromagnetically induced transparency (EIT) and Autler-Townes (AT) effect, including measuring the frequency \cite{hu2022continuously}, phase \cite{xie2025atomic, cai2023high, berweger2023closed}, amplitude \cite{hao2024microwave, sedlacek2012microwave}, polarization \cite{elgee2025electrically, you2024rf, cloutman2024polarization} and angle-of-arrival \cite{robinson2021angle, robinson2021determining} of the MW field. Recent studies mainly focus on improving the sensitivity and frequency response bandwidth. For instance, sensitivity improvements via superheterodyne  \cite{jing2020atomic}, critical point of many-body system  \cite{ding2022enhanced} and electric field enhancement  \cite{liu2025cavity, liang2025cavity, peng2018cavity, meyer2021waveguide, holloway2022rydberg}, bandwidth enhancement through Stark effect \cite{song2024continuous, liu2022highly,li2023super}, Zeeman effect \cite{fan2024microwave, shi2023tunable, akhmedzhanov2007electromagnetically}, multi-channel excitations \cite{hu2023improvement}, MW frequency combs \cite{zhang2024tunable, zhang2022rydberg} and space division multiplexing \cite{zhang2024ultra}, MW-photon conversion based on multi-wave mixing \cite{kasza2025atomic, borowka2024continuous, tu2022high, firdoshi2022six, vogt2019efficient}, etc. To date, the state-of-the-art Rydberg receivers can achieve a record sensitivity of $12.5~\rm{nV cm^{-1} Hz^{-1/2}}$ \cite{cai2022sensitivity} and a maximum instantaneous bandwidth of 54.6~MHz \cite{yan2025multi}. These breakthroughs have advanced the application scope of Rydberg receivers, including wireless communication \cite{yuan2024rydberg, liu2022deep, nowosielski2024warm}, MW vector determination \cite{sedlacek2013atom, richardson2025study}, dark matter detection \cite{graham2024rydberg}, and quantum thermometry \cite{schlossberger2025primary}. {\color{black} For communication systems, such high sensitivity enables a lower system noise floor, which enhances the signal-to-noise ratio (SNR) without requiring an increase in transmitter power. This improvement directly enhances the data transmission robustness and extends the capability for reliable detection of weak signals. Meanwhile, the pursuit of broader bandwidth addresses the fundamental need for greater frequency agility and higher spectral utilization density in advanced spectral-spread techniques like FHSS.}

    The inherent multi-frequency characteristics of FHSS systems exhibit natural compatibility with the ultra-broadband features of Rydberg receivers. {\color{black} Integrating Rydberg receivers into FHSS systems is expected to reduce the system complexity via eliminating the need for dedicated, wideband electronic mixers and their associated high-power LO drivers, while enhancing the sensitivity and the anti-jamming capability.} Recent advances have proposed and preliminarily verified the concept of Rydberg atom-enabled multi-frequency communication. For example, Zou et al. \cite{zou2020atomic} verified the feasibility of the frequency division multiplexing based on Rydberg atoms in wireless communication with bit error rates (BER) lower than 5$\%$ and data transfer rate of 200 kbps, utilizing amplitude modulation and frequency modulation. Du et al. \cite{du2022realization} achieved two-channel communication via coupling a single initial state to different final states, while communication frequencies were constrained on certain frequency bands as the initial state is fixed. \textcolor{black}{Wen et al. \cite{wen2024rydberg} demonstrated 20000 hop/s} frequency-hopping communication based on coherent population trapping, with a tunable bandwidth of 50 MHz, but limited to dual-frequency hopping. Recently, Chen et al. \cite{chen2025new, chen2025radar} developed a 1 GHz bandwidth system using non-uniform stepped-frequency synthesis, yet requiring complex laser frequency tuning. However, no experimental realization has achieved truly arbitrary-frequency hopping across a broadband using Rydberg atoms. This outstanding challenge underscores the critical need for an ultra-broadband arbitrary-frequency reception with a single Rydberg initial state.

    In this work, we introduce a waveguide-coupled Rydberg receiver for ultra-wideband FHSS communications and demonstrate its exceptional anti-jamming capability. {\color{black}Rydberg atoms serve as an intrinsic wideband mixer that simultaneously receives both the LO and signal fields, enabling direct down-conversion across the entire operational band with high sensitivity and lower LO power requirements compared to conventional RF mixers.} Our Rydberg atom-based FHSS system experimentally {\color{black}accomplishes} direct Binary Phase Shift Keying (BPSK) modulation and demodulation across an ultra-broad operational bandwidth of 100 kHz to 20 GHz (spanning 17 octaves from P to K band). It achieves a symbol rate of 100 kbps with {\color{black}maximum hopping rate of 100 khop/s} and supports over 150,000 channels. {\color{revblue} Furthermore, it achieves a channel density of approximately 8 channels per MHz within the 2.4 GHz ISM band. Bluetooth provides a useful reference point here, as its parameters are both well-defined and widely recognized. The Core Specification defines a system operating within a 79 MHz bandwidth at 1,600 hops per second across 79 channels, corresponding to a channel density of approximately 1 channel per MHz. This familiar benchmark helps illustrate the scale of channel diversity that ultra-broadband Rydberg receivers can potentially offer.} Based on communication anti-jamming theory, we quantitatively evaluate the superior anti-jamming capability of our system utilizing the jamming-resistance factor ($C_{\rm{JR}}$). Image transmission experiments {\color{black} demonstrate that it maintains} a BER consistently below $1\%$ under 0 dBm interference, representing outstanding robustness and a significant improvement in comparison with conventional systems. Crucially, experimental results prove that it supports all standard modulation types including amplitude, frequency and phase modulation, indicating potential for diverse ultra-broadband scenarios. The properties of Rydberg atoms are utilized to achieve direct signal demodulation, which simplifies the communication process while reducing system complexity.
    {\color{black}
    Our work establishes the first fully functional waveguide-enhanced Rydberg atomic receiver system for ultra-broadband FHSS secure communications without modifications to the optical configuration, thereby validating a practical framework for future high-performance anti-jamming communication.
    }

\section*{Results}
    \begin{figure*}
        \centering
        \includegraphics[width=1\linewidth]{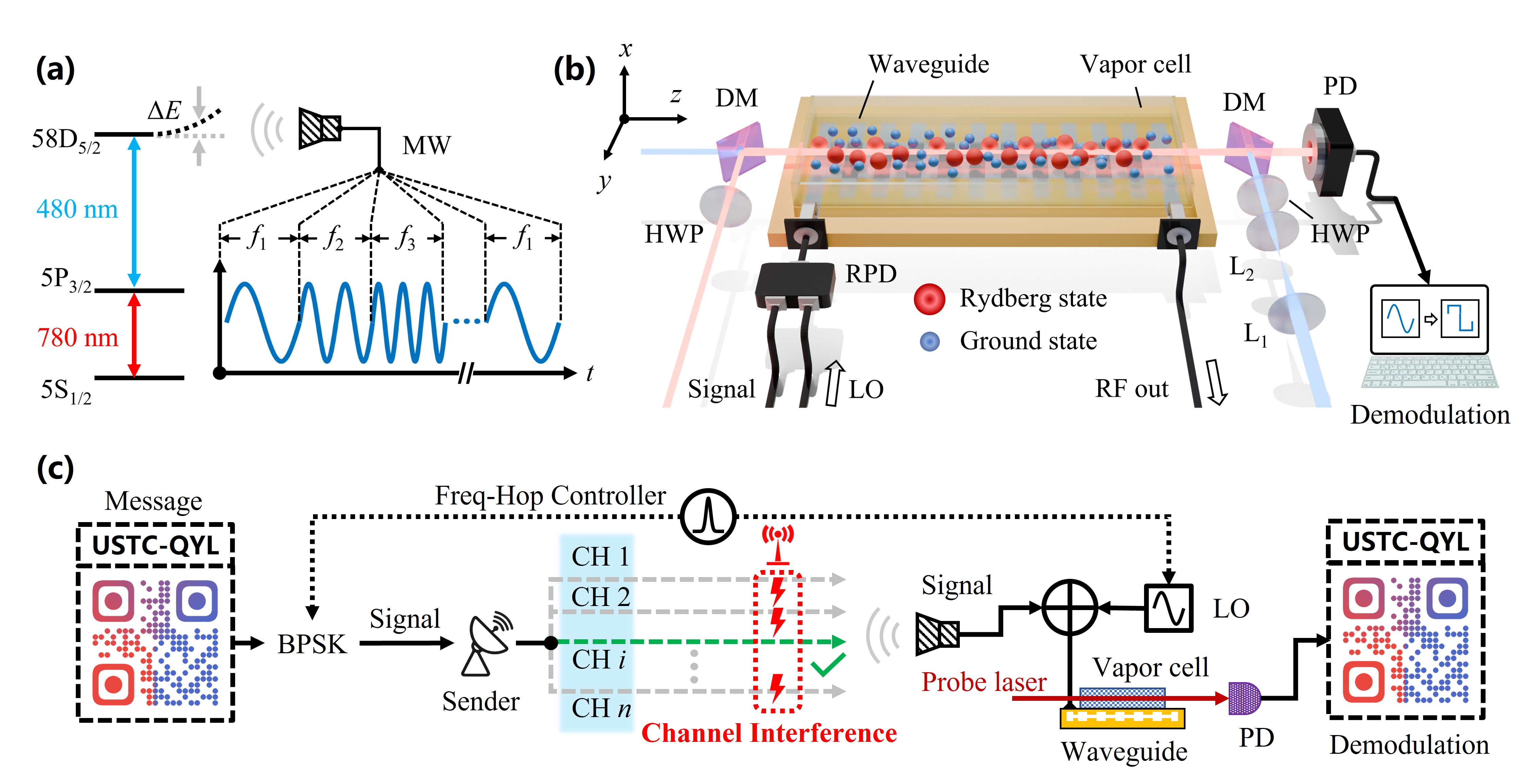}
        \caption{\textbf{Illustration of the Rydberg atom-based FHSS system.} (a) Energy diagram of the $^{85}\rm{Rb}$ atom. \textcolor{black}{The probe ($780\rm{~nm}$ laser, beam waist ${400\,\mu\text{m}}$) and coupling ($480\rm{~nm}$ laser, beam waist ${410\,\mu\text{m}}$) laser} couple the atomic states of the ground state $5\rm{S}_{1/2}$, the intermediate state $5\rm{P}_{3/2}$ and the Rydberg state $58\rm{D}_{5/2}$, realizing a configuration of electromagnetically induced transparency (EIT). In the presence of an MW field, the Rydberg state $58\rm{D}_{5/2}$ exhibits an energy change $\Delta E$ depending on the AC Stark effect. The frequency of the MW field is allowed to vary continuously or periodically within a certain range. (b) Schematic of the experimental setup. MW signals and local oscillator (LO) field are combined within a resistance power divider (RPD) and input into the waveguide directly. {\color{black}Note that the RPD functions only to linearly combine the two fields, producing their algebraic sum rather than an intermediate frequency signal.} The probe and coupling lasers counter-propagate through the waveguide-coupled Rydberg vapor cell, then the EIT transmission signal of the probe laser is collected by a photodetector (PD). (c) Illustration of the ultra-wide frequency-hopping communication based on Rydberg atoms. The message is BPSK-modulated onto a microwave carrier with frequency hopping across multiple channels ($\rm{CH}_{1}-\rm{CH}_{\it{n}}$, highlighted in blue shading). While external interference may disrupt certain channels, error-free transmission is statistically guaranteed by exploiting massive channels ($n\sim10^{5}$ for example) and an ultra-broad hopping range (tens of GHz). The receiver recovers data via BPSK-demodulation enabled via synchronized frequency-hopping between the carrier and LO fields, where the sparse distribution of jammed channels ensures communication reliability.}
        \label{fig1}
    \end{figure*}
    
    \begin{figure*}
        \centering
        \includegraphics[width=1\linewidth]{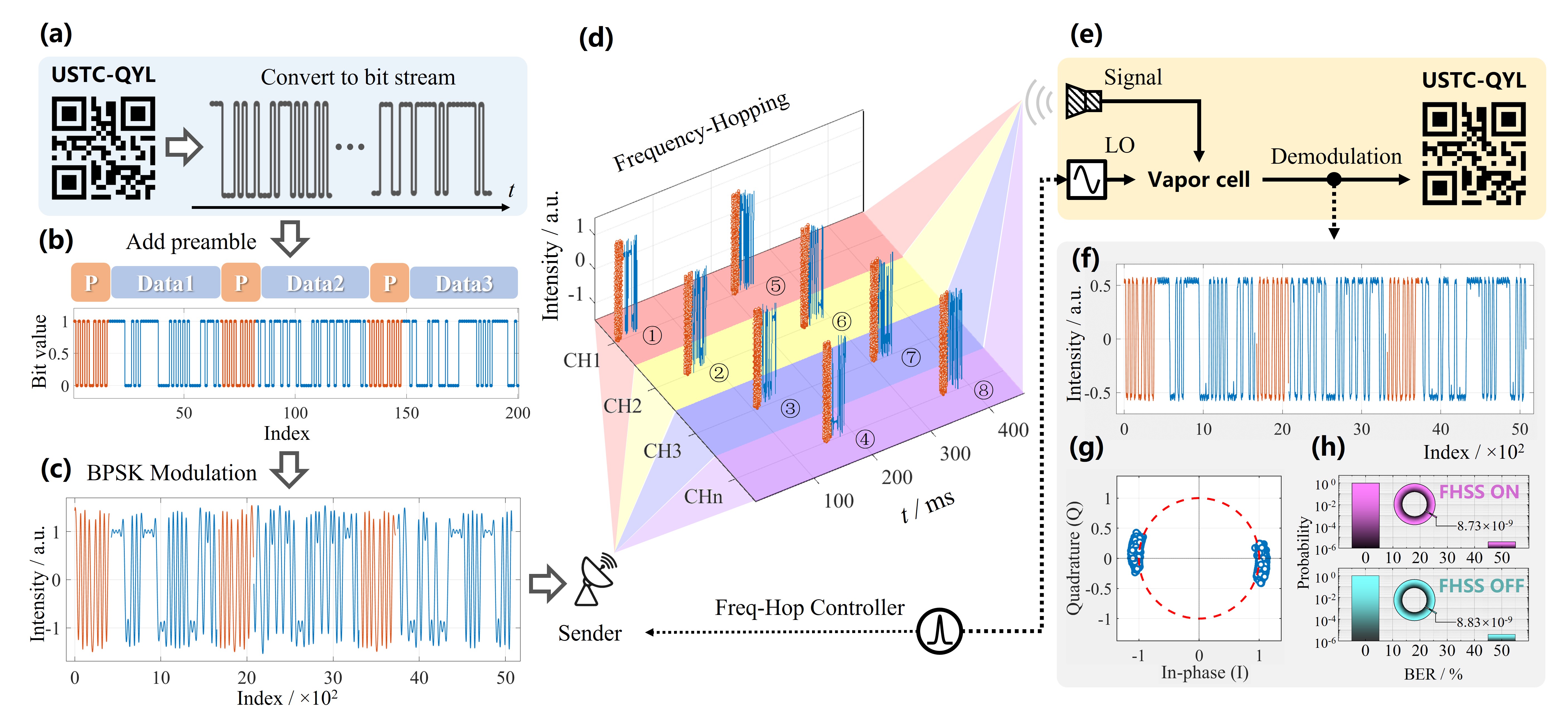}
        \caption{\textbf{Flow chart of the signal transmission and reception.} The massage (using a designated QR code as an example, shown in (a)) is first converted to a bit stream. The stream is then segmented into several data pieces, as illustrated in (b), each prefixed with a specific 16-bit preamble (fixed as 1010101001010101 in our implementation). These data pieces are BPSK-modulated onto a MW carrier at 25 samples/bit (shown in (c)), then transmitted with the carrier frequency pseudo-randomly hopped across the color-coded channels ($\rm{CH}_1-\rm{CH}\it{_n}$ in (d); note channel numbers are identifiers only). The signal is ultimately received (shown in (e)) by the Rydberg receiver via superheterodyne detection. Then the original massage can be recovered after demodulation. Figure (f) presents the stabilized intermediate frequency (IF) signal output from Rydberg atoms following automatic frequency and phase locking. The quality of the reception signal is further characterized in figure (g) and (h). (g) Constellation diagram of the atomic IF signal, where cluster compactness directly correlates with the signal fidelity. (h) Frequency histograms of bit error rate (BER) performance for the FHSS-enhanced (FHSS ON, magenta) and single-frequency (FHSS OFF, cyan) systems, evaluated over 259,643 and 256,833 test datasets (441 bits each), respectively. Both systems achieve quasi-error-free transmission, with average BERs $< 8.73\times10^{-9}$ (FHSS ON) and $< 8.83\times10^{-9}$ (FHSS OFF). The isolated 50$\%$ BER bars in each histogram originate from the final truncated cycle in time-limited tests, instead of system errors. Inset pies: Proportion of correctly received bits versus erroneous bits.}
        \label{fig2}
    \end{figure*}

     \begin{figure*}
        \centering
        \includegraphics[width=1\linewidth]{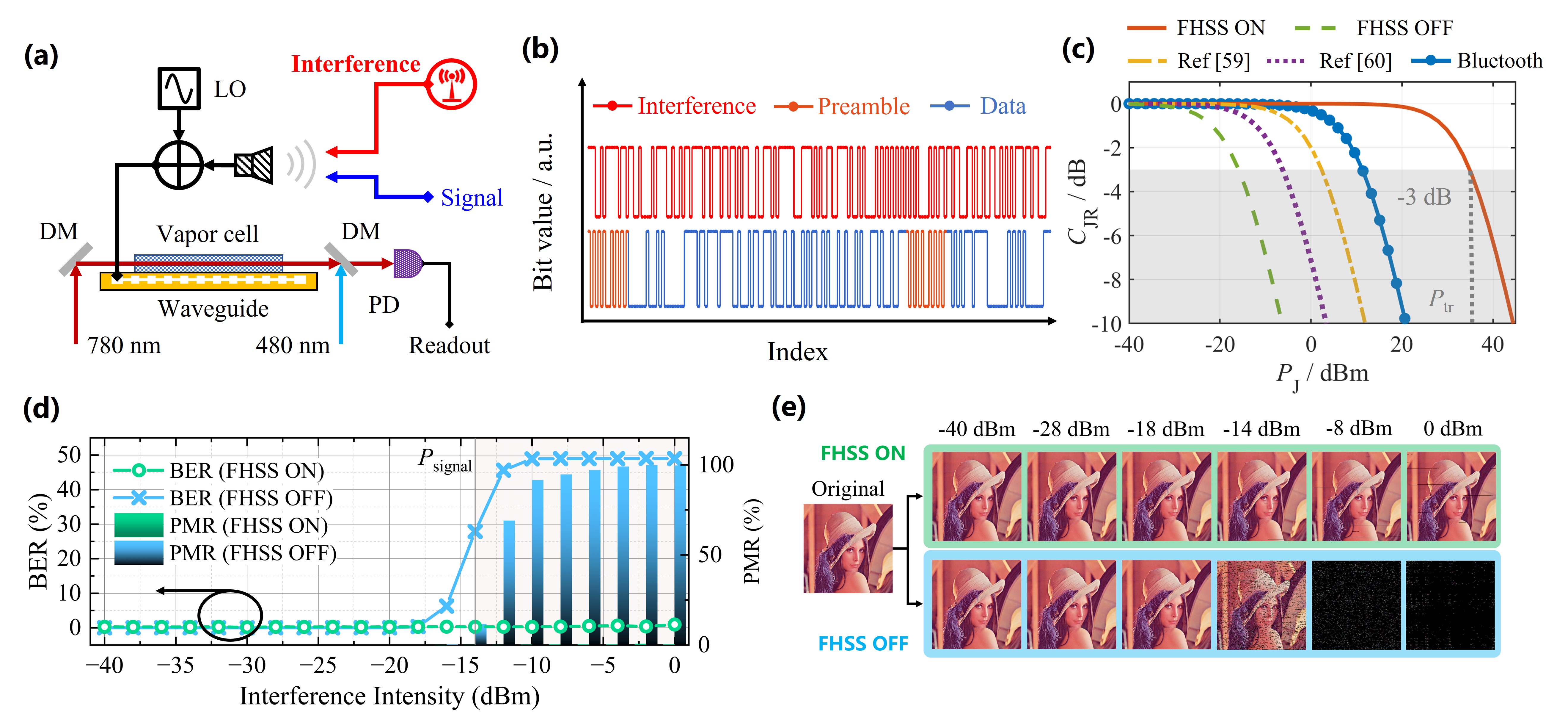}
        \caption{\textbf{Experimental results on the interference resistance of the frequency-hopping.} (a) Schematic diagram of the frequency-hopping Rydberg receiver under external interference. (b) Comparison of bit values between the interference and the signal. The empirically applied interference comprises randomly generated bit sequences (red dots and line). To evaluate the interference rejection capability of the frequency-hopping Rydberg receiver during signal reception, the jamming signal is deliberately configured to match the symbol rate (100 kbps) and carrier frequency (1.5 GHz) of the signal. (c) Jamming resistance capability ($C_{\rm{JR}}$) comparison between our FHSS-enhanced system (FHSS ON), a single-frequency system (FHSS OFF), 
        {\color{black}
        a conventional Bluetooth system \cite{muller2000bluetooth}
        }
        and dual-frequency hopping systems (using methodologies from reference \cite{du2022realization} and \cite{wen2024rydberg}). Systems with $C_{\rm{JR}}$ below -3 dB (gray-shaded region) exhibit significant interference susceptibility, indicating that our Rydberg atom-based FHSS system exhibits outstanding anti-jamming performance. (d) Measured Bit error rate (BER) and pixel missing rate (PMR) based on statistical analysis of {\color{black}64,000} test datasets versus the interference intensity for systems with (green) and without (blue) FHSS implementation. The system without FHSS implementation exhibits rapid BER/PMR degradation when the interference exceeds the signal intensity (approximately -14 dBm), whereas the FHSS-enhanced Rydberg receiver maintains stable performance {\color{black}(BER $< 1\%$, PMR $< 1\%$)} across all tested interference levels. (e) Image transmission quality comparison under varying interference conditions, demonstrating the robustness of the system with FHSS implementation (green background) versus the system without FHSS implementation (blue background). These results conclusively validate the immunity of the FHSS-enhanced Rydberg receiver to single-tone jamming, with at least a 16 dBm improvement in interference tolerance.}
        \label{fig3}
    \end{figure*}

    \begin{figure*}
        \centering
        \includegraphics[width=1\linewidth]{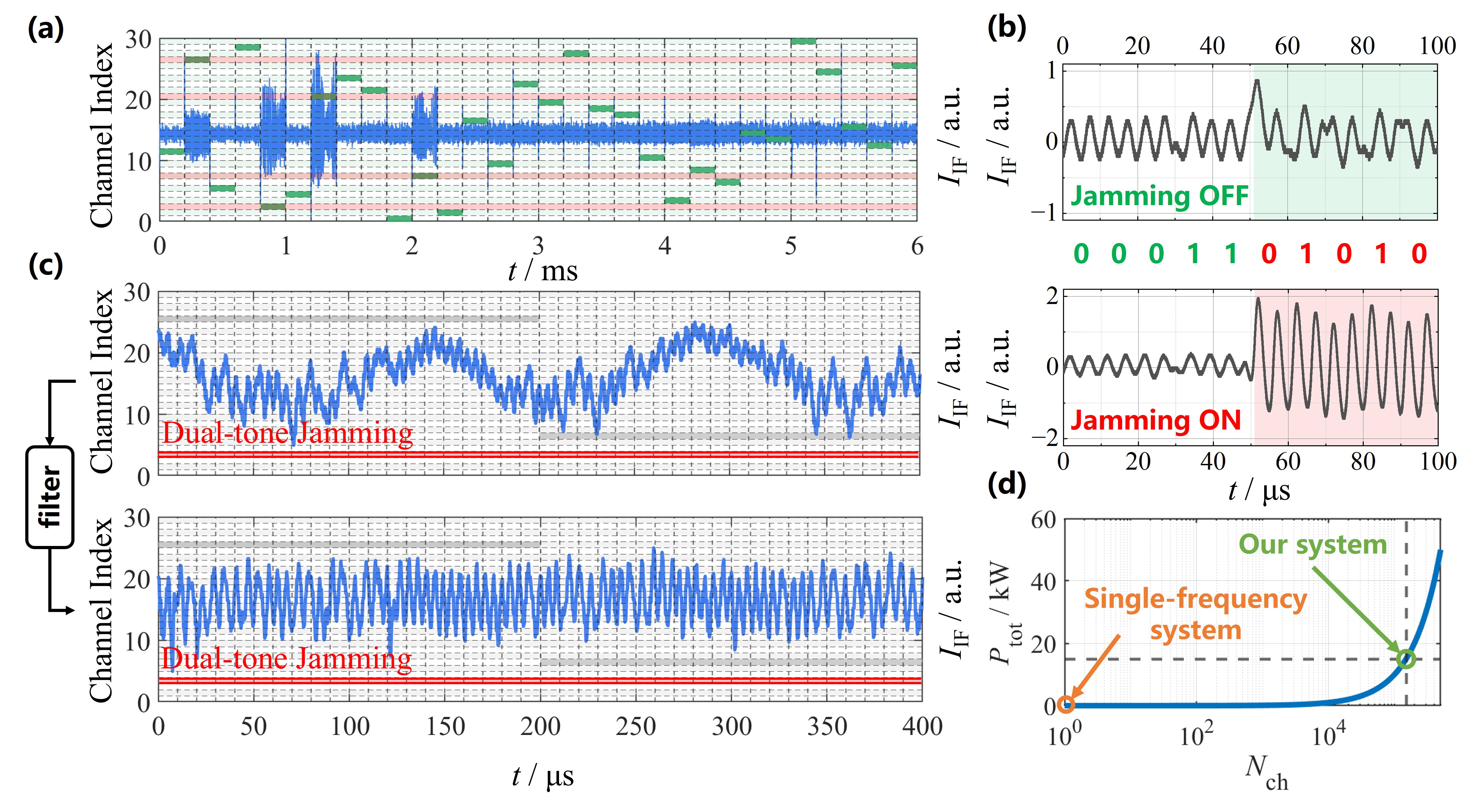}
        \caption{\textbf{Response of the Rydberg atom-based FHSS system under multi-frequency interference.} (a) Rydberg atomic IF output signals (blue traces) under four independent narrowband jammers. The background shading alternates between light green and white stripes to distinguish communication channels, with green blocks indicating the active hopping channel (i.e., the hopping pattern) and red stripes marking jammed channels. (b) Comparison between the transmitted bit stream and the atomic IF output signals. In interference-free channels (green background in upper panel), the IF signals accurately decode the transmitted bits, whereas strong interference (red background in lower panel) corrupts the IF signals integrity, devastating information recovery accuracy. (c) The atomic IF signal (solid blue line) under a dual-tone jamming with a 10 kHz frequency offset (denoted by red double lines), where the gray blocks represent the active hopping channels. The upper panel reveals the low-frequency modulation arising from the beat note between the two jammers, while the lower panel demonstrates complete data recovery after bandpass filtering, confirming effective interference rejection. (d) Total jamming power required ($P_{\rm{tot}}$) to suppress all channels versus the number of system channels ($N_{\rm{ch}}$). The power requirement increases prohibitively with $N_{\rm{ch}}$, highlighting the interference-resistant advantage of our Rydberg atom-based FHSS system.}
        \label{fig3_1}
    \end{figure*}
    
    \begin{figure*}
        \centering
        \includegraphics[width=1\linewidth]{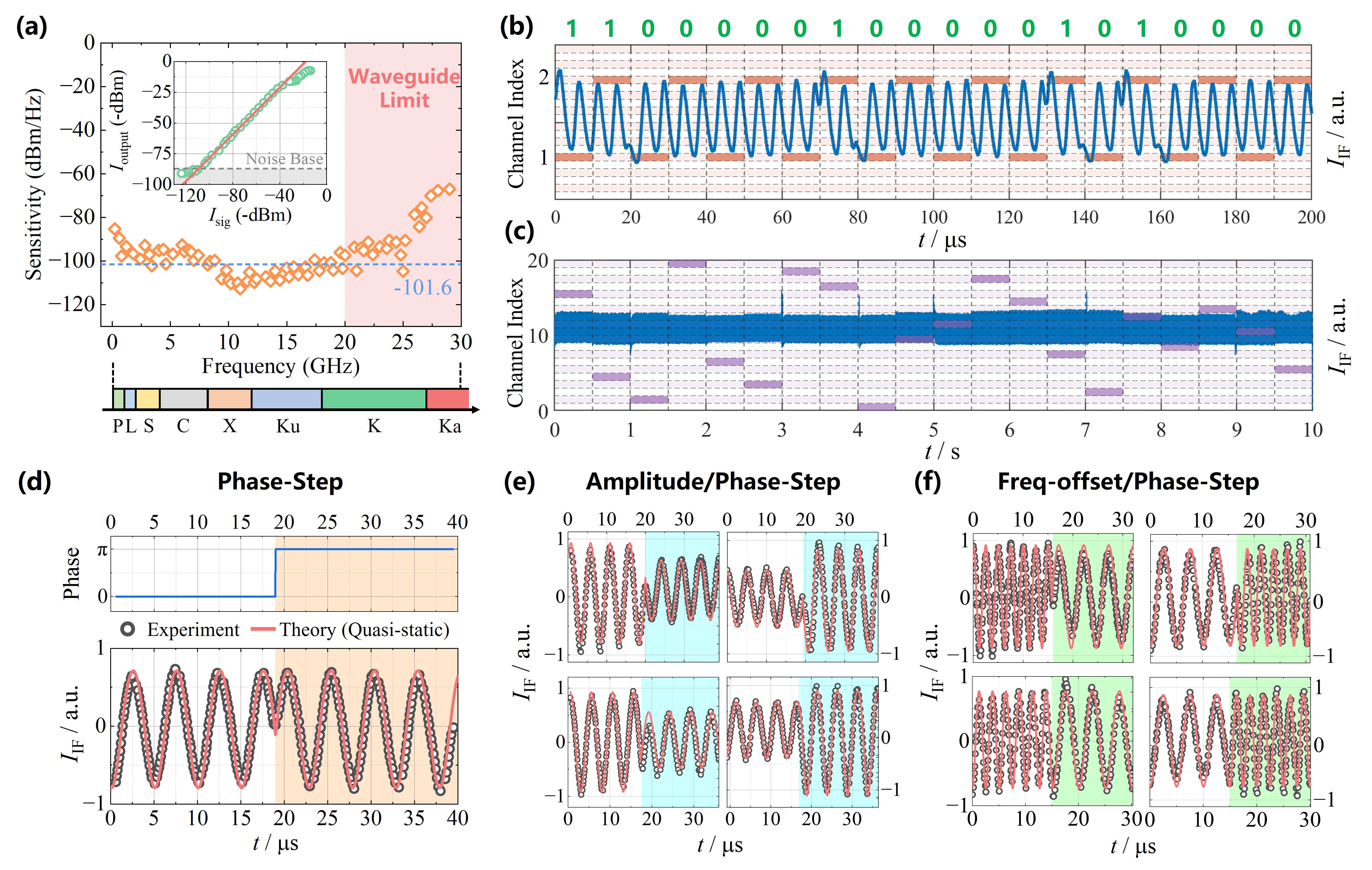}
        \caption{\textbf{Ultra-broadband reception and instantaneous frequency-hopping response.} (a) Intrinsic sensitivity of the Rydberg receiver (orange squares) after a correction of the insertion loss shows a frequency-independent response with an average value of -101.6 dBm/Hz (blue dashed line). It exhibits a degradation beyond 20 GHz due to the waveguide coupling efficiency limitations. The inset displays the output IF intensity versus the input signal intensity at 10.77 GHz, where the -87 dBm/Hz noise floor yields a sensitivity of -109.3 dBm/Hz. (b) Ultra-fast dual-channel frequency-hopping communication. Constrained by the instantaneous bandwidth (132 kHz) and the symbol rate (100 kbps) of the Rydberg receiver, a minimum dwell time of 10 $\mu$s per channel (corresponding to a hopping rate of 100 khop/s) is achieved at two selected frequencies (10.7 GHz and 11.7 GHz, orange blocks). The decoded bit sequence obtained from the IF signal (blue curve) under a fixed frequency offset of 200 kHz is displayed above the panel. (c) Ultra-broadband multi-channel frequency-hopping communication. The LO frequency follows a pseudo-random hopping sequence (purple blocks) within an operational bandwidth spanning 100 kHz to 20 GHz (divided into 20 predefined channels), while maintaining a 200 kHz frequency offset. The corresponding atomic IF output (blue curve) demonstrates stable operation under a symbol rate of 100 kbps and a dwell time of 500~ms. (d)$\sim$(f) Instantaneous response of Rydberg atoms to step-function perturbations. (d) Phase-step ($0\rightarrow\pi$ at $t = 19~\rm{\mu s}$, post-step data on orange background), (e) Amplitude-steps ($\pm 5$ dBm) with/without $\pi$ phase-steps (post-step data on blue background), (f) Frequency-offset steps (200 kHz $\leftrightarrow $ 400 kHz) with/without $\pi$ phase steps (post-step data on green background). Experimental data (black circles) and theoretical predictions (red curves) show excellent agreement across all tested configurations.}
        \label{fig4}
    \end{figure*}
    
    \textbf{Ultra-broadband Communication Model}
    A complete communication system comprises four fundamental components: the transmitter, the channel, the receiver and the information protocol. Under ideal transceiver assumptions and given that the external interference is uncorrelated with the spreading codes (e.g., when the jammer cannot acquire the spreading sequence), the spectrum-spread communication system can randomize and spread the narrowband interference (with a total power of $P_{\rm{J}}$) across the entire operational bandwidth $B_{\rm{op}}$. Consequently, the effective interference power passing through the receiver filter (with bandwidth $B_{\rm{D}}$) is reduced by a processing gain $G_{\rm{P}}=B_{\rm{op}}/B_{\rm{D}}$. To quantify the anti-jamming capability of the systems, we define a jamming-resistance factor $C_{\rm{JR}}$ as follows\cite{pickholtz2003theory}:
     \begin{align}
        C_{\rm{JR}}=\frac{\rm{SJR}}{\rm{SNR}}=\frac{\eta_{0}}{\eta_{0}+\frac{P_{\rm{J}}}{B_{\rm{op}}}}
        \label{eq1}
     \end{align}
    where $\rm{SNR} = \it{P}_{\rm{s}}/ \left ( \eta_{\rm{0}} B_{\rm{D}} \right )$ is the interference-free signal-to-noise ratio, $\rm{SJR}=\it{P}_{\rm{s}}/ \left ( \it{P}_{\rm{J}}/\it{G}_{\rm{P}}+\eta_{\rm{0}}B_{\rm{D}} \right )$ denotes the actual signal-to-jamming ratio, $P_{\rm{s}}$ represents the signal power and $\eta_{0}$ is the power density of the background white noise. Notably, the metric $C_{\rm{JR}} \in \left (0, 1 \right )$ quantifies the ability to recover information with jamming present. A value approaching unity indicating stronger anti-jamming capability, while a value of zero signifies complete suppression by the interference. The operational bandwidth of the system determines its resilience against malicious interference. Therefore, increasing the operational bandwidth $B_{\rm{op}}$ in a FHSS communication system can significantly enhance its anti-jamming capability. 

    In Rydberg atom-based communication systems, the operational bandwidth is determined by the maximum number of channels $N_{\rm{ch}}$:
    \begin{align}
         B_{\rm{op}}=N_{\rm{ch}}\times B_{\rm{ins}}
         \label{eq2}
    \end{align}
    where $B_{\rm{ins}}$ is the instantaneous bandwidth, referring to the maximum frequency range that the receiver can process simultaneously. The energy diagram of the Rydberg atoms is presented in Fig.~\ref{fig1}(a). When exposed to MW fields with varying frequencies, the Rydberg states experience AC Stark shifts that alter the transmission spectrum of the probe laser. These spectral variations are directly detected by a photodetector (PD), with the corresponding experimental setup detailed in Fig.~\ref{fig1}(b) (see Method section). The schematic diagram of the Rydberg atom-based ultra-broadband FHSS system is demonstrated in Fig.~\ref{fig1}(c), where the transmission and recovery of the string “USTC-QYL” under external interference is exhibited. As any frequency component within the operational bandwidth can be arbitrarily selected, the maximum number of allowed channels is expressed as $ N_{\rm{ch}}^{\rm{max}} = B_{\rm{op}}/B_{\rm{ins}} $. 
    {\color{black}
    Notably, the frequency-hopping control process illustrated in Fig.~\ref{fig1}(c) is indispensable for realistic deployments featuring physically separated signal sender and receiver. In such scenarios, high-precision synchronization between them can be achieved using well-established techniques. A cost-effective and widely deployed method utilizes Global Navigation Satellite System (GNSS, e.g., GPS) disciplined oscillators, which provide synchronization at the nanosecond level \cite{costa2004performance}. For applications requiring even higher precision, fiber-based time-frequency distribution techniques offer an alternative with accuracy up to tens of picoseconds \cite{schnatz2014neat, li2025high}. This enables a remote LO filed to precisely follow the hopping sequence of the sender via distributed clock and timing signals. In our proof-of-principle experiments, for simplicity, both the signal and LO fields are generated by a single co-located instrument (see Method).
    }

    \textbf{Frequency-Hopping Communication}
    The flow chart in Fig.~\ref{fig2} illustrates the frequency-hopping communication process in the Rydberg atom-based FHSS system. Assuming the information to be transmitted is the string “USTC-QYL”, we first convert it into a $21\times21$ Quick Response (QR) code, which is subsequently transformed into an initial bit stream (441 bits in total) consisting of binary symbols (0 and 1) as shown in Fig.~\ref{fig2}(a). To ensure reliable carrier-frequency identification at the Rydberg receiver, the initial bit stream is segmented with each data piece prefixed by an identical 16-bit binary preamble (1010101001010101), yielding the reformatted bit stream shown in Fig.~\ref{fig2}(b). Before signal transmission, the reformatted bit stream is modulated onto a carrier wave utilizing the BPSK method, as shown in Fig.~\ref{fig2}(c), with the carrier frequencies continuously hopping under the control of a frequency-hopping controller. By assigning different carrier frequencies to distinct channels ($\rm{CH}_{1}...\rm{CH}_{\it{n}}$ in Fig.~\ref{fig2}(d)), the transmitted signal undergoes rapid channel switching during propagation. When subjected to external malicious interference, the Rydberg atom-based FHSS system demonstrates an enhanced anti-jamming performance, benefiting from its exceptionally broad hopping bandwidth. In our experimental setup, both the local oscillator (LO) field and the signal carrier hop frequencies synchronously following a predetermined pseudo-random frequency-hopping sequence, while preserving a constant frequency offset. \textcolor{black}{By simultaneously receiving both fields, the Rydberg atoms transcribe all the information from the signal field into the beat note between the LO and the signal, which manifests in the transmission spectrum of the probe laser and is detected by a PD (the coupling laser and other optical components are omitted for clarity)}. Finally, as illustrated in Fig.~\ref{fig2}(e), the original transmitted string is recovered by performing BPSK demodulation on the PD output.

    We first determine the instantaneous bandwidth and dynamic range of the Rydberg receiver, which determine the upper limit of distinguishable signals and the lower threshold of detectable signal strength, respectively. The instantaneous bandwidth, referring to the maximum frequency range that the receiver can process at a given moment, is measured to be 132 kHz by fixing the LO frequency at 1.5 GHz while varying the detuning $\rm{\delta} \it{f}$ between the LO and the signal fields. Notably, the system achieves an unprecedented frequency-hopping range spanning 100 kHz to 20 GHz (20 GHz operational bandwidth), yielding a maximum of 151,515 available hopping channels. This represents a 1,900-fold improvement over conventional Bluetooth systems in mobile devices (79 channels across 79 MHz bandwidth) \cite{muller2000bluetooth} and surpasses previous records \cite{du2022realization, wen2024rydberg} (2-channel systems with $\leq 10$ MHz bandwidth) by four orders of magnitude. As evidenced by Eq.~(\ref{eq1}), our system demonstrates overwhelming advantages in both frequency adaptability and channel diversity, providing quantum-enhanced anti-jamming capabilities.

    The dynamic range, referring to the difference in signal strength that the receiver can simultaneously handle, is quantified using a spectrum analyzer (Ceyear, 4024F) in 1~Hz resolution bandwidth. The total dynamic range of the Rydberg receiver reaches approximately 102~dB. Both the instantaneous bandwidth and the dynamic range are ultimately constrained by the -128.03~dBm noise floor from the optical read-out. While the dynamic range achieved here is not the cutting-edge due to the limited coupling laser power, it can be further enhanced through improving Rydberg excitation efficiency with a higher coupling laser power or laser stabilization using an ultra-stable cavity to reduce the read-out noise. Additionally, the minimum detectable intensity of the MW field is determined to be -138.17~dBm after accounting for the 10.14~dB insertion loss from the RPD and signal transmission wires.

    \textbf{Quality of Signal Transmission}
    During the BPSK signal demodulation (Fig.~\ref{fig2}(e)), representative results in Fig.~\ref{fig2}(f) demonstrate successful information recovery at 25 samples/bit. Both the preamble prefixed in each data piece and the initial data embedded during the BPSK modulation can be clearly identified. The transmission quality of the Rydberg atom-based FHSS system is quantified via the constellation diagram shown in Fig.~\ref{fig2}(g), where the in-phase (I) and the quadrature (Q) components form distinct clusters at (±1, 0), exhibiting typical BPSK characteristics. Each point in the constellation diagram corresponds to a single bit, representing the amplitude and the phase distribution of the demodulated signal. The observed cluster compactness indicates a superior transmission quality as well as higher fidelity.
    
    As a key metric for communication robustness, the bit-error rate (BER) is statistically evaluated using 259,643 (FHSS-enabled) and 256,832 (single-frequency) datasets, with each containing 441 bits. The histogram in Fig.~\ref{fig2}(h) shows perfect error-free transmission (BER = 0, magenta bar) for both the FHSS-enhanced system (upper panel) and the single-frequency system (lower panel), corresponding to an average BER lower than $8.73\times10^{-9}$ and $8.83\times10^{-9}$ respectively. Both the systems exhibit ultra-low BER and demonstrate high communication quality and resistance to bit-loss/inversion errors under the interference-free condition. It is noteworthy that this performance is achieved without error-correction algorithms. Implementation of advanced error-correction coding schemes (e.g., LDPC coding \cite{richardson2002design}) could further enhance the BER of the Rydberg atom-based FHSS system.
    
    \textbf{Anti-Jamming Capability}
    The anti-jamming performance of the Rydberg atom-based FHSS communication system is experimentally quantified in Fig.~\ref{fig3}. As schematically shown in Fig.~\ref{fig3}(a), a co-channel interference with the same frequency as the carrier of the desired signal is introduced into the Rydberg receiver. To stimulate challenging interference conditions, we employed the BPSK-modulated interference with random bit sequences synchronized to the symbol rate of the signal. Representative bit sequences are displayed in Fig.~\ref{fig3}(b), where red, orange, and blue traces correspond to interference signals, preamble and data packets respectively. The jamming-resistance $C_{\rm{JR}}$ quantifies the SNR degradation under specified interference power, with $C_{\rm{JR}}$ down to -3~dB defining the interference tolerance $P_{\rm{tr}}$ where SJR reduces to half of the interference-free SNR. Figure.~\ref{fig3}(c) compares the jamming-resistance factor $C_{\rm{JR}}$ across four Rydberg atom-based systems versus conventional Bluetooth. The background noise density $\eta_{0}$ in Eq.~(\ref{eq1}) is calibrated to -67.7 dBm/Hz via the measured interference power tolerance $P_{\rm{tr}}$ of the single-frequency systems. While the dual-frequency hopping systems (yellow dash-dotted and purple dotted) exhibit marginal improvement in  $P_{\rm{tr}}$ over the system without FHSS implementation, their performance remains inferior to that of Bluetooth systems. In contrast, our FHSS-enhanced system demonstrates significant improvement, with $P_{\rm{tr}}$ enhancements of 51 dBm and 23 dBm relative to the system without FHSS implementation and Bluetooth systems, respectively. These quantitative results confirm the exceptional interference resistance of our proposed system. 
    
    Quantitative measurements in Fig.~\ref{fig3}(d) reveal that the system without FHSS implementation (blue curves) suffers from rapid degradation in both the BER and the pixel missing rate (PMR) when the interference intensity exceeds the signal power ($P_{\rm{sig}} = -14$ $\rm{dBm}$). In contrast, the FHSS-enhanced system (green curves) maintains an average BER and PMR {\color{black}below $1\%$}, respectively, across all tested interference levels. We attribute these residual errors primarily to limitations in the data processing algorithm, with Fig.~\ref{fig2}(h) demonstrating that further algorithmic optimization can achieve even lower error rates. This remarkable performance is visually confirmed in the image transmission tests (Fig.~\ref{fig3}(e)), where the image integrity from the FHSS-enhanced system (top row, green background) contrasts sharply with the severely distorted results from the system without FHSS implementation (bottom row, blue background) under strong interference. 

    The anti-jamming performance of the Rydberg atom-based FHSS system is systematically evaluated under multiple interference conditions, with experimental results exhibited in Fig.~\ref{fig3_1}. {\color{black}The experiment employs a representative set of thirty randomly selected frequencies spanning 1–3 GHz (numbered 1–30). This setup creates a demanding anti-jamming scenario where four specified channels (1.14, 1.88, 2.40, and 2.76 GHz) are subjected to interference, corresponding to a high jammed-channel fraction of $13\%$.} Figure.~\ref{fig3_1}(a) presents the frequency hopping sequence and corresponding atomic IF outputs, with jammed channels marked in red stripes. In interference-free regions (green background, Fig.~\ref{fig3_1}(b)), the system maintains stable IF signals that enable accurate bit decoding, whereas channels subject to a 10 dBm single-tone interference (red background) exhibit severe signal distortion that prevents data recovery. {\color{black}
    The results confirm that interference remains confined to the targeted channels, while all others maintain full communication integrity. This validates the multi-channel operability of the system and highlights its inherent scalability, as a fixed number of jammer channels would cause proportionally less degradation as the total available channel set expands.
    }
    {\color{black}
    The inherent robustness of the system is supported by its vast theoretical channel pool (over 150,000 available channels, see supplementary materials). This extensive resource directly translates into supreme frequency agility, thereby ensuring reliable operation even when multiple channels experience localized interference.
    }
    Additional low-frequency modulation is observed in the IF signals when two jammers with 10 kHz frequency offset coexist in a single channel, as depicted in Fig.~\ref{fig3_1}(c). This distortion is effectively mitigated using a 200 kHz bandpass filter matched to the frequency-offset between the LO and signal fields. Remarkably, even under more challenging conditions where multiple modulated interferers occupy a single channel, the original data still can be successfully recovered utilizing advanced filtering algorithms.

    In our experiments, the minimum jamming power needed for an effective single-channel interference ($P_{\rm{sg}}$) is approximately 10 dBm under optimal operating conditions with a peak SNR, at which point the IF outputs of the Rydberg receiver are completely distorted. It implies that full-spectrum suppression for multi-channel systems requires a total jamming power of $P_{\rm{tot}} = P_{\rm{sg}} \times N_{\rm{ch}}$, as demonstrated in Fig.~\ref{fig3_1}(d). Compared to single-frequency systems, complete jamming of our system demands prohibitively higher power due to its ultra-large channel capacity, rendering it fundamentally impractical to achieve a full-spectrum suppression. These results validate the exceptional interference resilience and outstanding practicability of our Rydberg atom-based FHSS system. 

   \textbf{Ultra-broad Frequency-Hopping Bandwidth}
    Experimental results demonstrate that the Rydberg receiver is capable of detecting MW across a frequency range of 100 kHz to 20~GHz. To analyze the spectrum performance, its intrinsic sensitivity (with insertion loss correction) is characterized across a broad range extending to 30~GHz (Fig.~\ref{fig4}(a)), maintaining a fixed 200~kHz frequency offset $\delta f$ between the LO and signal fields. Limited by the available microwave source bandwidth and waveguide coupling efficiency, the frequency-hopping demonstrations are specifically conducted within the 100 kHz to 20 GHz range. The insert displays a output-input power relationship of the receiver at a LO frequency of 10.77 GHz, where the gray region denotes the noise base of the MW field measured in 1 s integration time, indicating a sensitivity of -80.9~dBm/Hz. The ultra-broad reception bandwidth (equivalent to the frequency-hopping bandwidth) spans 17 octaves, covering the P, L, S, C, X, Ku and K bands. It is noteworthy that the current frequency-hopping bandwidth can be further extended by employing a MW source with broader frequency coverage. 
    
    Figure~\ref{fig4}(b) illustrates the fastest frequency-hopping communication process achievable under current device limitations. To ensure the RF source stability while maintaining high hopping performance, we implement a dual-frequency hopping scheme between 10.7 GHz and 11.7 GHz. The resulting atomic IF signal (blue curve) reveals that each binary symbol (0 or 1) is represented by two complete waveform cycles, demonstrating a frequency-hopping rate of 100 khop/s with single-symbol transmission per channel per hop.
    {\color{black}
    We note that the IF signals from both channels in Fig.~\ref{fig4}(b) share the same 200 kHz frequency, which makes them indistinguishable from the IF alone. To further validate this high hopping rate, we therefore conducted an additional experiment leveraging the narrowband selectivity of Rydberg atoms to directly resolve the temporal hopping sequence of the fields, with results providing direct proof of the 100 khop/s hopping dynamics (see supplementary materials).} 
    To demonstrate the broadband processing capability of the Rydberg receiver, we segment the operational bandwidth (100 kHz to 20 GHz) into 20 discrete channels (100 kHz-1 GHz, 1-2 GHz, ..., 19-20 GHz). The carrier frequency of the signal field randomly switches among these channels with a dwell time of 500 ms per channel. Crucially, the carrier frequency selection is randomized such that even when revisiting the same channel at different times, distinct frequency values may be employed. Throughout the experiments, the symbol rate of the signal field is maintained at 100 kbps, with a sampling rate of 25 samples/bit and a 200 kHz frequency offset between the LO and signal fields. The corresponding IF signal output from the Rydberg atoms is presented in Fig.~\ref{fig4}(c).

    \textbf{Instantaneous Response in Frequency-Hopping}
    We characterize the instantaneous response of Rydberg atoms to frequency-hopping MW signals with an oscilloscope directly, as shown in Fig.~\ref{fig4}(d)$\sim$(f). When maintaining the same amplitude as well as frequency offset $\delta f$ between the LO and the signal fields before/after frequency-hopping, the variation of the atomic IF output before/after the $\pi$ phase-step is recorded in Fig.~\ref{fig4}(d). The upper panel shows the phase transition for the signal field from 0 to $\pi$ (post-step period highlighted in orange), while the lower panel presents the corresponding IF signal dynamics (black circles) alongside quasi-static theoretical predictions (red curve; see Methods), showing excellent agreement. Further measurements under simultaneous amplitude/phase-steps and frequency-offset/phase-steps (Fig.~\ref{fig4}(e)$\sim$(f), post-transition data in blue and green) demonstrate the robust tolerance of the Rydberg receiver to abrupt MW field variations in intensity, frequency, and phase. These results indicate that our BPSK-optimized Rydberg receiver can be extended to various modulation schemes, highlighting its potential for advanced communication applications.

\section*{Discussion}

    As previously demonstrated in Fig.~\ref{fig2}, the Rydberg receiver achieves successful demodulation of BPSK-modulated signals and information recovery of FHSS transmission. 
    {\color{revblue}Our Rydberg atom-based FHSS system demonstrates an ultra-broadband operational bandwidth spanning 100 kHz to 20 GHz with a hopping rate of 100 khop/s. As a reference, established frequency hopping systems such as Bluetooth operate within 79 MHz at 1600 hop/s \cite{muller2000bluetooth} while FHSS WiFi systems use 79 MHz at 4500 hop/s \cite{huang2002scalability}. Notably, when operating within the same frequency band such as the 2.4 GHz ISM band, our system achieves a channel density approximately eight times higher than that of Bluetooth and meets the sensitivity requirement of -70 dBm. These results illustrate that Rydberg atom-based sensors can achieve performance levels consistent with established communication protocols under identical spectral constraints.}
    This exceptional performance stems from the intrinsic broadband response of Rydberg atoms, suggesting that the operational bandwidth could be further extended by employing more advanced microwave sources and broadband waveguides. The Rydberg receiver also achieves excellent interference immunity through the advantages of supporting more than 150,000 channels, ultra-broad bandwidth operation, and ultra-fast hopping capability. Quantitative analysis shows 32 dBm and 51 dBm improvements in narrowband interference tolerance compared to conventional single-frequency systems and previous reported works\cite{du2022realization, wen2024rydberg}, respectively.
    {\color{black}
    Furthermore, experimental verification confirms that over $91\%$ of channels remain available even under severe resonant jamming (see Supplementary Materials).
    }
    These advancements originate from our innovative detection methodology. It utilizes the AC Stark effect and heterodyne detection, circumventing the limitations of MW-field resonant interactions inherent in traditional approaches.

    Notably, the maximum symbol rate of our system is currently constrained to approximately 100 kbps due to the instantaneous bandwidth limitation of the Rydberg receiver, which consequently restricts the achievable hopping rate. 
    {\color{black}
    Although the current laser beam configuration is sufficient for our experiments, the instantaneous bandwidth could be further enhanced through technical optimizations such as reducing the beam waist and increasing the laser power \cite{yang2024highly, yan2025multi}. A broader bandwidth would directly translate to higher achievable symbol and frequency-hopping rates, suggesting a strengthened jamming resistance of the Rydberg atom-based FHSS system.
    }
    It should be emphasized that the theoretical relaxation time for Rydberg atoms to reach a steady-state under step-signal excitation is {\color{black}approximately 1 $\mu$s (see Method)}, fundamentally enabling a maximum achievable {\color{black}frequency-hopping rate of 1 Mhops/s for the current system.
    }
    {\color{black} However, this is not a fundamental limit of the Rydberg receiver concept. Recent work in optimized systems has demonstrated relaxation times down to 150 ns \cite{prajapati2024investigation}, suggesting the potential for future implementations to support hopping rates exceeding 6 Mhop/s. Thus, our demonstration validates the principle of ultra-broadband frequency-hopping reception, while pointing to a substantial performance headroom achievable with further engineering of the excitation and microwave source parameters.
    }

    {\color{black}
    Furthermore, regarding practical implementation, the role of the signal-capturing front-end warrants further consideration. In any real-world implementation aiming to exploit the whole 100 kHz – 20 GHz operational bandwidth demonstrated here, a suitable antenna (or set of antennas) is required to couple incident microwave into the system (as shown in Fig.~\ref{fig1} (c)). However, due to the fundamental bandwidth-size constraint for antennas (often described by the Chu-Harrington limit \cite{bing2008emerging}), it is extremely challenging for a single antenna to cover the entire range from kHz to tens of GHz. A practical solution to leverage the exceptional operational bandwidth of the system would be to employ a set of antennas, each optimized for a specific sub-band. This front-end complexity is inherent to any ultra-broadband system aiming to enhance communication security through expanding the spectrum diversity. 
    {\color{revblue}The main advancement in this work lies in drastically simplifying the receiver chain at the back. Due to the fundamental physical limitations of electronic mixing, a conventional receiver covering the same frequency range would require multiple parallel RF chains where each includes its own oscillator, mixer, and filters. In contrast, our Rydberg atom-based system functions as a single unified back end that replaces this entire hardware bank while offering software defined frequency agility without reconfiguration.}
    {\color{revblue}The optical components used in our current demonstration including lasers and the vapor cell are amenable to miniaturization through established techniques. Recent advancements in micromachined vapor cells and integrated photonic platforms suggest a potential path toward fully integrated portable devices.\cite{Anderson2024}} 
    Thus, while the antenna front-end represents a common challenge for all ultra‑broadband schemes, the Rydberg approach achieves a radical reduction in back‑end complexity and tuning overhead. Future improvements in atomic sensitivity may eventually enable direct coupling of electric field to Rydberg atoms. However, even with current antenna technology, the integration of a Rydberg back‑end signifies a major step toward simpler, more adaptive broadband communication and sensing systems.
    }

    {\color{black}
    In summary, we have demonstrated the first integrated ultra-broadband anti‑jamming communication system based on a waveguide‑coupled Rydberg atomic receiver. Departing from previous resonant approaches, our off‑resonant heterodyne scheme enhanced by a waveguide overcomes the inherent trade‑off between bandwidth and sensitivity, enabling continuous operation from 100 kHz to 20 GHz (spanning P through K bands, 17 octaves) in a frequency-hopping spread spectrum configuration with a hopping rate up to 100 khop/s. 
    {\color{revblue}The system performance was evaluated in the context of established standards such as Bluetooth, demonstrating wider bandwidth, higher channel density of approximately 8 channels per MHz, a faster frequency hopping rate, and stronger inherent interference immunity.}
    This immunity is intuitively and quantitatively verified through image‑transmission experiments that maintain a 100 kbps symbol rate with high fidelity (BER $<$ 1 $\%$) under jamming intensities up to 0 dBm. {\color{revblue}The 1 $\%$ threshold serves as a common reference in anti-jamming performance evaluations, reflecting the robustness of the receiver under stressed conditions\cite{ETSI_EN_300_113}. }Although the current demonstration is constrained by available hardware (notably microwave sources), the intrinsic broadband nature of Rydberg atoms indicates that the fundamental capabilities far exceed the reported metrics. These point toward potential hopping rates approaching $\sim$1 Mhop/s and operational bandwidths spanning an even broader continuous spectrum (e.g., from near-DC to $>$20 GHz). This work establishes a practical framework for Rydberg‑atom‑based secure communications and highlights a clear path toward their implementation in jamming‑resistant networks, quantum‑enhanced sensing, and related quantum‑technology applications.
    }

\section*{Method}

    \textbf{Experimental Setup}
    As the energy level diagram illustrated in Fig.~\ref{fig1}(a), the Rubidium atoms are coherently excited from the ground state $5\rm{S}_{1/2}$ to the Rydberg state $58\rm{D}_{5/2} $ via a two-photon EIT process. The fundamental working mechanism of our Rydberg atomic receiver relies on the (off-resonant) AC Stark effect, which mediates the effective coupling between the MW fields and Rydberg atoms. Specifically, when the Rydberg atoms interact with synchronized frequency-hopping LO and signal MW fields, they exhibit a time-dependent AC Stark shift $\Delta E$ that tracks the predetermined frequency-hopping sequence. Consequently, the Rydberg receiver achieves an ultra-broad operational bandwidth due to the off-resonant coupling with the MW field through the AC Stark effect. 
    
    Figure.~\ref{fig1}(b) illustrates the experimental configuration of the ultra-broadband frequency-hopping Rydberg receiver. 
    {\color{black}
    The LO and signal MW fields are generated and synchronized using two instrument setups depending on the experiment. For the maximum hopping-rate demonstration (Fig.~\ref{fig4}(b)), we used a dual-channel vector signal generator (Rohde $\&$ Schwarz SMW200A). Both channels play synchronized hopping-sequence waveform files, with Channel B operating in a hardware‑triggered "Retrigger Baseband A" mode to ensure precise, nanosecond-level hopping transitions. For other frequency-hopping communication experiments, a software‑defined radio platform (Ettus Research USRP X310) is employed. Synchronization is achieved via centralized software control within GNU Radio, with both channels sharing a common clock and executing hops simultaneously based on coordinated FPGA timestamps. In both cases, the combined MW output is fed through a resistance power divider into a waveguide for field enhancement.
    }
    A pair of counter-propagating lasers, including the probe laser (Toptica, DLpro 780) and the coupling laser (Toptica, DLpro480), are precisely aligned using dichroic mirrors (DM) and focusing lenses to intersect at the Rydberg vapor cell positioned above the waveguide. The identical linear polarization of the beams is assured with half-wave plates (HWP). When excited to the Rydberg state ($\Omega_{\rm{p}}=2\rm{\pi}\times7.49 ~\rm{MHz}, \Omega_{\rm{c}}=2\rm{\pi}\times4.44 ~\rm{MHz}$), the atoms in the vapor cell interact with external MW fields containing both the free-space LO and the signal field. In our experiment, the beat note at frequency $\delta_{f}=\left | f_{\rm{LO}}-f_{\rm{sig}}\right |$, produced by the heterodyne mixing between the LO field $E_{\rm{LO}}$ and the signal field $E_{\rm{sig}}$, is extracted from the Rydberg EIT spectrum utilizing a Thorlabs PDA36A photodetector (PD), enabling subsequent demodulation to recover the original modulated information.

    \textbf{Off-resonant interaction in Rydberg vapor}
    \begin{figure}
        \centering
        \includegraphics[width=1\linewidth]{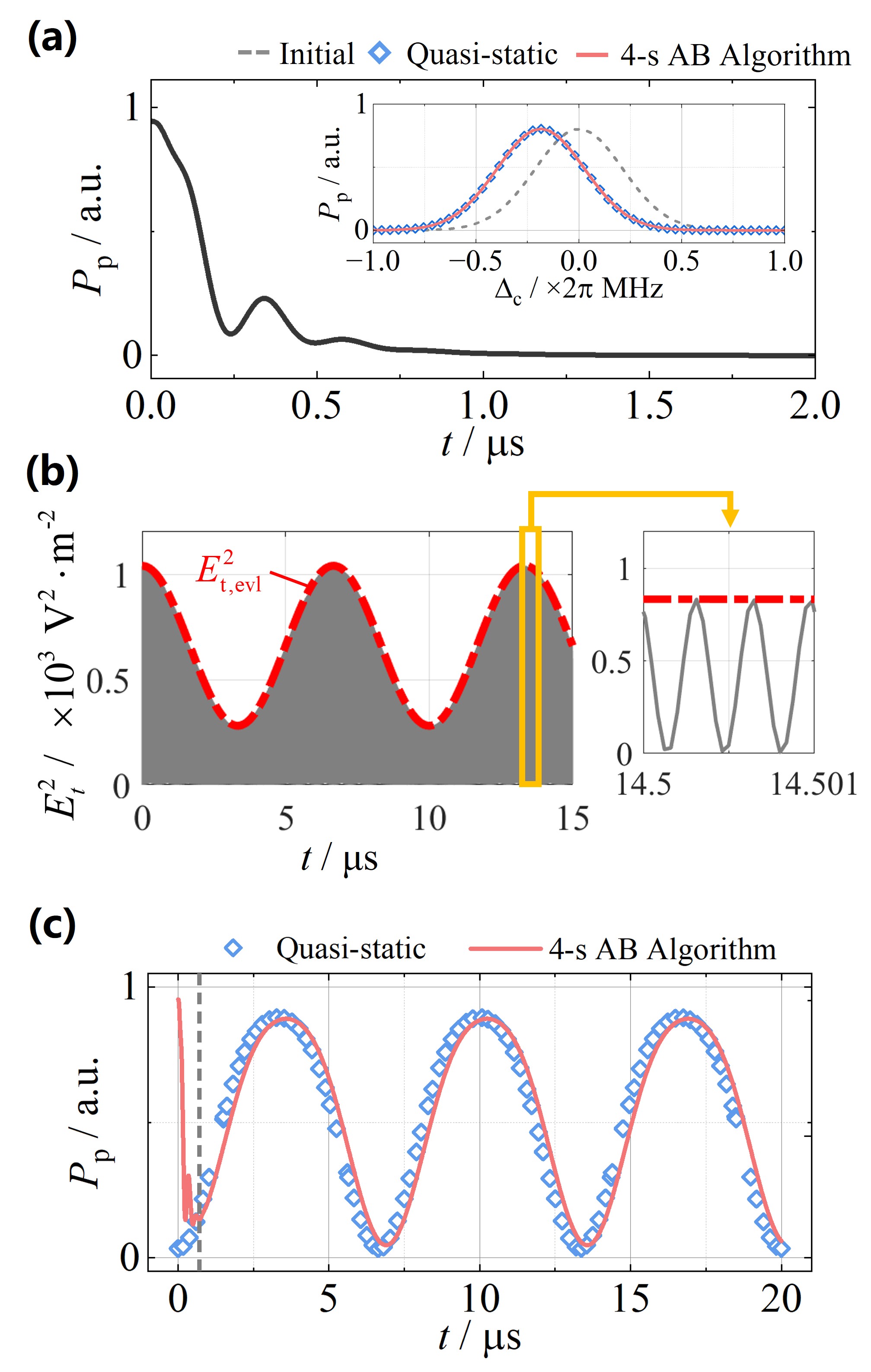}
        \caption{\textbf{Comparison between the quasi-static approximation and the four-step Adams-Bashforth (4-s AB) method.} (a) Initially prepared in the Rydberg state without an MW field present, the atoms reach a steady-state when irradiated by 1.5 GHz MW field, with the corresponding probe transmission shown as the black curve. The inset displays the steady-state EIT spectra obtained via the quasi-static method (blue squares) and the 4-s AB method (red curve), with the initial EIT spectrum shown as the gray dashed curve. (b) Correlation between the total electric field $E_{\mathrm{t}}^2$ and its envelope $E_{\mathrm{t,evl}}^2$ (right inset: 10 ns zoom). (c) The IF signal calculated through both methods with a 1.5 GHz LO field frequency detuned by $\delta f=200$ kHz.}
        \label{method_fig1}
    \end{figure}
    The off-resonant interaction between the Rydberg atoms and the external MW field is conventionally described quantum-mechanically using the density matrix $\rho$. The temporal evolution of the atomic state is governed by the Lindblad equation:
    \begin{align}
        \frac{\partial\rho}{\partial t} = \frac{1}{ \rm{i}\hbar } \it{ \left[ \hat{H}, \rho \right] + \hat{\mathcal{L}}\left[ \rho \right]}
        \label{eq3}
    \end{align}
    where $\hat{H}$ denotes the Hamiltonian of the system and $\hat{\mathcal{L}}\left [ \cdot \right ]$ is the Lindblad super-operator accounting for atomic decoherence (e.g., spontaneous emission). 
    {\color{black}
    To accurately simulate the transient response dynamics and predict the relaxation time, our model incorporates an additional dark-state-correction \cite{bohaichuk2022origins}. Here the modified Hamiltonian in the absence of the microwave field is expressed as:
    \begin{align}
        \hat{H}_{0} = \frac{\hbar}{2}
        \begin{bmatrix}
          0 & \Omega_{\rm{p}} & 0 & 0 \\
          \Omega_{\rm{p}} & -2\Delta_{\rm{p}} & \Omega_{\rm{c}} & 0 \\
          0 & \Omega_{\rm{c}} & -2\left(\Delta_{\rm{p}}+\Delta_{\rm{c}}\right) & 0 \\
          0 & 0 & 0 & 0
        \end{bmatrix}
    \end{align}
    Here, $\hat{H}_0$ denotes the Hamiltonian of the atomic system coupled to the optical fields prior to the introduction of the microwave perturbation. $\Delta_{\rm{p}}$ and $\Delta_{\rm{c}}$ represent the detunings of the probe and coupling lasers from their respective atomic resonances, $\Omega_{\rm{p}}$ and $\Omega_{\rm{c}}$ are the corresponding Rabi frequencies, and $\hbar$ is the reduced Planck constant. The fourth basis state corresponds to a metastable dark state, which accounts for the extended coherence time observed in the transient response.
    }
    Since the applied MW field is non-resonant with any Rydberg transitions in our experimental setup, its effect on the atomic states should be characterized through the AC Stark shift. For the combined LO and signal WM fields, the resulting Stark shift reads:
    \begin{align}
        \Delta E= -\frac{1}{2} \alpha\left [(E_{\rm{LO}}\left( t \right)+E_{\rm{sig}}\left( t \right) \right ]^{2}
        \label{eq4}
    \end{align}
    \begin{align}
        \alpha_{0}=2e^{2} \sum_{n'l'j' \neq nlj} \frac{\left| \left< n,l,j,m_{j} \left| r_{0} \right| n',l',j',m'_{j} \right> \right|^{2}}{E_{n'l'j'}-E_{nlj}}
        \label{eq5}
    \end{align}
    with $ E_{\rm{LO}}$ and $ E_{\rm{sig}}$ being the respective electric field amplitudes, $E_{nlj}$ is the energy of the state $\left| n,l,j\right>$, and $\alpha$ denotes the Rydberg-state polarizability calculated utilizing the ARC (Alkali Rydberg Calculator) python package\cite{vsibalic2017arc}. We employ the time-dependent Lindblad equation as the basis for the parametric simulation used as an aid to interpret the experimental results we present. To facilitate the verification of our theory, the density matrix $\rho\left( t \right)$ obtained from the Lindblad equation is converted into the transmission of the probe laser $P\left( t \right)$. 
    \begin{align}
        P_{\rm{p}}\left( t \right)=P_{\rm{in}} \rm{e}^{-\it{k}_{\rm{p}} \it{L}_{\rm{cell}} \rm{Im}\left[\chi\left(t\right)\right]}
        \label{eq6}
    \end{align} 
    \begin{align}
        \chi\left( t \right)=-\frac{2 N_{0} \mu_{\rm{p}}^{2}}{\varepsilon_{0} \hbar \Omega_{\rm{p}}}\rho_{21}\left(t\right)
        \label{eq7}
    \end{align}

    In our model, the atoms are initially prepared in the ground state $\left | 5\rm{S}_{1/2} \right >$ and allowed to evolve freely without the MW field until reaching a steady state. The Rydberg atom exhibits a static Stark shift under monochromatic MW irradiation while demonstrating negligible response to high-frequency (GHz-scale, for example) MW field (Fig.~\ref{method_fig1}(a)). These findings validate the omission of high-frequency terms in Eq.~(\ref{eq4}) and the solution of the Lindblad equation can consequently be simplified utilizing a quasi-static approximation. Specifically:
    \begin{equation}
        E_{\rm{t}}^{2}  =  \left[ E_{\rm{LO}}\left( t \right) + E_{\rm{sig}}\left( t \right) \right]^{2}
    \end{equation}
    \begin{equation}
        E_{\rm{t,evl}}^{2} = 
                 2  E_{\rm{LO}}E_{\rm{sig}}\cos\left ( \delta \omega t +\delta\varphi \right )  + \left( E_{\rm{LO}}^{2}+E_{\rm{sig}}^{2}\right)
    \end{equation}
    where $E_{\rm{LO}}\left( t \right) = E_{\rm{LO}} \cos\left( \omega_{\rm{LO}}t + \varphi_{\rm{LO}} \right)$ and $E_{\rm{sig}}\left( t \right) = E_{\rm{sig}} \cos \left( \omega_{\rm{sig}}t + \varphi_{\rm{sig}} \right)$ are the electric components of the LO and signal MW field respectively, $\delta \omega = \omega_{\rm{sig}}-\omega_{\rm{LO}}$ and $\delta \varphi = \varphi_{\rm{sig}}-\varphi_{\rm{LO}}$ represent their frequency and phase offset. Fig.~\ref{method_fig1}(b) compares the total electric field intensity $E_{\rm{t}}^{2}$ with its envelope $E_{\rm{t,evl}}^{2}$. Since $E_{\rm{t,evl}}^{2}$ represents the instantaneous value of $E_{\rm{t}}^{2}$, the effective field $E_{\mathrm{t,eff}}^{2} = E_{\mathrm{t,evl}}^{2}/2$ is adopted in the quasi-static approach, following the dependence of AC Stark shift on the root-mean-square (RMS) value of the MW electric field. 
    {\color{black}
    It should be emphasized that the quasi-static approach described above, which directly loads the intermediate-frequency beat note onto the Rydberg atoms by filtering out the high-frequency components, represents a deliberate theoretical approximation adopted to conserve computational resources. While this approach does not fully replicate the true dynamic interaction process between the atoms and the combined LO–signal field, it captures the essential atomic response behavior by omitting finer dynamic details, thereby offering a computationally efficient alternative for system-level analysis and design. The more faithful description of actual atomic dynamics is provided by the transient solution method, which is implemented here using the four-step Adams-Bashforth (4-s AB) algorithm.
    } 
    As shown in Fig.~\ref{method_fig1}(c), the exact transient solution calculated via 4-s AB algorithm exhibits excellent agreement with the result of the quasi-static approximation.

    \begin{figure}
        \centering
        \color{black}
        \includegraphics[width=1\linewidth]{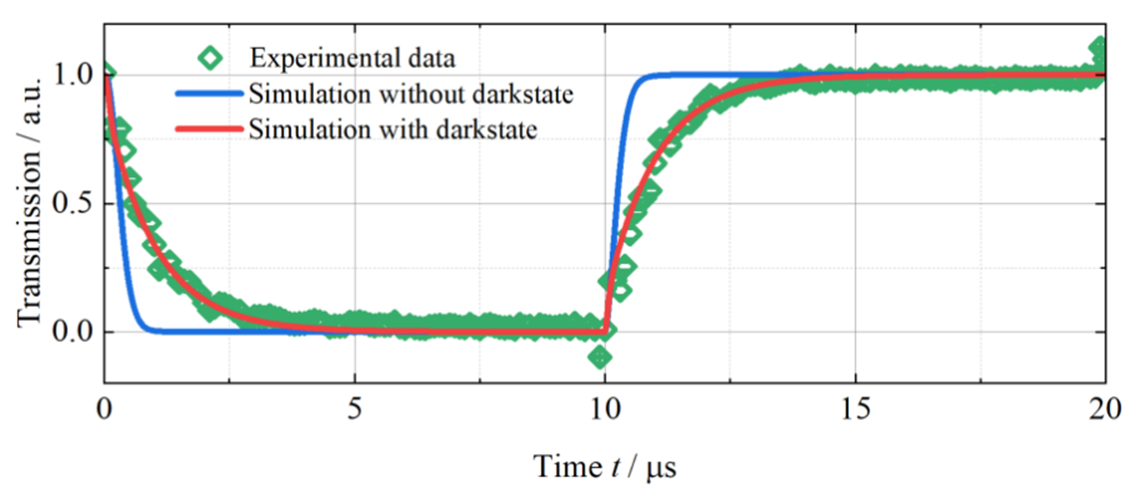}
        \caption{\textbf{The transient response of the system to a step-function MW field.} The dark-state correction is adopted to model our specific $^{85}\rm{Rb}$ Rydberg receiver system. The blue curve shows the prediction of the uncorrected model, with a relaxation time of 0.2 $\mu s$. The red curve shows the result of the dark-state-corrected model, predicting a relaxation time of 1 $\mu s$. The green diamonds represent the experimental data from five repeated runs (error bars indicate standard deviation), which show excellent quantitative agreement with the corrected theoretical prediction.}
        \label{method_fig2}
    \end{figure}

    {\color{black}
    To accurately assess the dynamic response limit of our Rydberg receiver, we incorporate a dark-state correction into our theoretical model to better capture its transient dynamics. As shown in Fig.~\ref{method_fig2}, when numerically solving the time-dependent Lindblad equation to simulate the instantaneous response of the system to a step-function electric field perturbation, the uncorrected model (blue curve) predicts a relaxation time of approximately 0.2 $\mu s$. In contrast, the dark-state-corrected model (red curve) yields a relaxation time of about 1 $\mu s$, which aligns quantitatively with our independent experimental measurements from five repeated runs (green diamonds; error bars indicate standard deviation). This corrected relaxation time corresponds to a theoretical maximum frequency-hopping rate of approximately 1 Mhop/s.

    It is noteworthy that this relaxation time ($\sim 1~\mu s$) is significantly shorter than the actual hopping interval ($\sim 10~\mu s$) achieved in our fastest frequency-hopping experiments. This temporal separation implies that the system has ample time to reach a steady state between successive frequency hops. Therefore, when analyzing the interaction with modulated signals, the use of a quasi-static approximation in theoretical calculations is well justified and does not compromise the physical accuracy.
    }

{\color{black}
\section*{DATA AVAILABILITY}
    The data generated in this study have been deposited in the Code Ocean database (\href{https://codeocean.com/capsule/7164889/tree}{https://codeocean.com/capsule/7164889/tree}).

\section*{CODE AVAILABILITY}
    The custom codes used to produce the results presented in this paper are available from the corresponding authors upon request.
}

\section*{Acknowledgements}
    The authors appreciate the instructive discussions with Prof. Ming-Min Zhao from Zhejiang University. We acknowledge funding from the National Key R and D Program of China (Grant No. 2022YFA1404002), the National Natural Science Foundation of China (Grant Nos. T2495253, 61525504, and 61435011).

\section*{Author contributions statement}
    D.-S.D., B.-B.W. and L.-H.Z conceived the idea. J.-D.N. and J.R.C conducted the physical experiments. The research was supervised by D.-S.D. and B.-B.W. All authors contributed to discussions regarding the results and the analysis contained in the manuscript.

\section*{Competing interests}
    The authors declare no competing interests.



%

\end{document}